\documentclass[format=acmsmall]{acmart}
  
\usepackage{amsmath}
\usepackage{caption}
\usepackage{color}
\usepackage{graphicx}
\usepackage{multirow}
\usepackage{subcaption}
\usepackage{url}

\newcommand{\ignore}[1]{}

\setcopyright{acmcopyright}
\copyrightyear{2020}
\acmYear{2020}
\acmDOI{XX.XXXX/XXXXXXX.XXXXXXX}

\acmJournal{TOIS}
\acmVolume{00}
\acmNumber{00}
\acmArticle{00}
\acmMonth{00}

\begin{document}

\title{Assessing top-$k$ preferences}

\author{Charles L. A. Clarke}
\email{claclark@gmail.com}
\author{Alexandra Vtyurina}
\email{sasha.vtyurina@gmail.com}
\author{Mark D. Smucker}
\email{mark.smucker@uwaterloo.ca}
\affiliation{%
  \institution{University of Waterloo}
  \city{Waterloo}
  \country{Canada}
}

\renewcommand{\shortauthors}{Clarke, Vtyurina and Smucker}

\begin{abstract}

Assessors make preference judgments faster and more consistently than
graded judgments.
Preference judgments can also recognize distinctions between items
that appear equivalent under graded judgments.
Unfortunately, preference judgments can require more than linear effort to
fully order a pool of items,
and evaluation measures for preference judgments are not as well established
as those for graded judgments,
such as NDCG.
In this paper,
we explore the assessment process for partial preference judgments,
with the aim of identifying and ordering the top items in the pool,
rather than fully ordering the entire pool.
To measure the performance of a ranker,
we compare its output to this preferred ordering by applying a rank
similarity measure.
%% We demonstrate the practical feasibility of this approach by crowdsourcing
%% partial preferences for the TREC 2019 Conversational Assistance Track,
%% replacing NDCG with a new measure that can reflect factors beyond relevance.
%% This new measure has its most striking impact when comparing modern neural
%% rankers,
%% where NDCG can fail to recognize significant differences exposed by this
%% new measure.
We demonstrate the practical feasibility of this approach by crowdsourcing
partial preferences for the TREC 2019 Conversational Assistance Track,
replacing NDCG with a new measure named \textit{compatibility}.
This new measure has its most striking impact when comparing modern neural
rankers,
where it is able to recognize significant improvements in quality that
would otherwise be missed by NDCG.

\end{abstract}

\maketitle




\section{Introduction}

Preference judgments~\cite{cbcd08,yao95,rorvig90,fs91,sz20}
have long been proposed as an alternative to graded judgments
for the offline evaluation of search and related ranking tasks,
including recommendation and question answering.
Instead of independently judging individual items according
to defined criteria,
assessors make preference judgments on pairs of items by comparing them
side-by-side to determine the better of the two.
If we allow ties, 
where  assessors are not forced to strictly prefer one item over the other,
preference judgments impose a weak ordering on a set of items.
To evaluate the performance of a ranker on a query,
we can directly compare this weak ordering to the actual ranking
generated for that query.
If we employ a rank similarity measure for this comparison,
it provides a measure of the ranker's performance~\cite{compat}.
This approach contrasts with the more established approach of converting
independently assigned grades into gain values to compute measures
such as NDCG~\cite{kk02,burges10} and ERR~\cite{cmzg09}.

Compared with independent graded judgments,
assessors make preference judgments faster and more
consistently~\cite{cbcd08}.
Preference judgments also allow assessors to make
assessments that better agree with actual user click preferences, and
it is hypothesized that the better agreement is the result of
a contextualization of the search results caused by viewing two
documents at once~\cite{kyct13}.
Preference judgments also make it easy to incorporate factors beyond
those captured by commonly used ordinal relevance scales~\cite{beyond,rel}.
For example, for e-commerce search,
these factors might include price and quality.
For a news search vertical,
these factors might include recency,
so that an assessor comparing two news stories of equal topical relevance, could choose
the latest update.
If two news articles have equal topical relevance and timeliness,
an assessor might prefer a shorter, more focused,
article over a longer article containing extraneous information.
Preference judgments can also take personalization into account,
so that locally available items could be preferred for an e-commerce search,
or concordant political views could be preferred for a news search.

Preference judgments face two criticisms.
First, preference judgments may not be transitive.
If item A is preferred over item B, and item B is preferred over item C,
we may not be able to assume that A would be preferred over item C.
If we can't assume transitivity,
a set of $n$ items may require $O(n^2)$ preference judgments.
Even if we assume transitivity,
a set of $n$ items requires $O(n\log{n})$ judgments to produce a total order.
In contrast, if we have dedicated and reliable assessors,
traditional graded assessment requires exactly $n$ judgments.
Second,
while NDCG and similar graded measures are well established
for offline evaluation in both industry and academia,
widely accepted evaluation measures for preference judgments have not yet
emerged~\cite{cbcd08,ymt18}.

In prior work,
we addressed the first criticism by proposing
\emph{evaluation by partial preferences}~\cite{compat}.
We focus preference judgments on identifying
and carefully ordering the best items for a query,
perhaps no more than four or five.
Since these are the items that are most likely to be seen by a
searcher~\cite{granka04,mustafa19},
these are the items a ranker should return as the top results,
ranked consistently with preferences.
These will have the most impact on perceived search quality,
and it's important to get them right.
The remaining items can be grouped into larger equivalence classes,
exactly as they are for graded measures,
so that they still contribute to the measurement of ranker performance,
but with less impact than the best items.

To address the second criticism, we measure a system's performance
by its \emph{maximum similarity to an ideal ranking}~\cite{beyond}.
Partial preferences impose a weak ordering on a collection.
We interpret this weak ordering as a set of ideal rankings for a query.
For the best items,
preference judgments can precisely define this ideal ordering.
For the larger equivalence classes,
any ordering of the items in the class is equally good,
although we do not include the class of non-relevant items in our ideal
rankings.
We then apply a rank similarity measure to compare these ideal rankings
to an actual ranking generated by the system we wish to measure.
As our performance measure,
we take the maximum similarity between the members of the ideal set and the
actual ranking.

We call this process of computing maximum similarity to a set of
ideal rankings computing the \emph{compatibility} of the actual ranking.
When compared to traditional graded measures,
compatibility allows us to more precisely specify the ideal response expected
from a ranker,
and to compare this ideal response with its actual response.
We provide further details regarding compatibility in
Section~\ref{sec:compat}.
As part of computing compatibility,
we use Rank Biased Overlap~\cite{wmz10} (RBO)
to compute similarity between ideal and actual rankings.
The properties of RBO make it ideally suited for this purpose,
and we provide further details regarding RBO in Section~\ref{sec:rbo}.

This thread of research~\cite{compat,beyond}
was directly motivated by our experience implementing offline evaluation
metrics for a social media site.
Even under carefully composed assessment guidelines,
multiple items may appear to be perfect,
but when these items are placed side-by-side,
a clearly desirable ordering becomes apparent.
For example,
on social media sites popular entertainers may have multiple official accounts.
As well, there may be multiple high quality and carefully curated
fan accounts.
At the time of writing,
there are at least two verified accounts for Taylor Swift on Twitter,
@taylorswift13 with 86M followers and @taylornation13 with 1M followers.
As well, there are multiple fan accounts with over 100K followers.
When independently assessed,
and seen outside the context of the others,
any of these accounts could reasonably be labeled as perfect for the query
``taylor swift''.
When placed side-by-side,
and considering factors such as the number of followers,
we might rank @taylorswift13 first,
@taylornation13 second,
with the various fan accounts after that.

Maximum similarity to an ideal ranking represents a radical simplification of
existing offline evaluation practice.
Essentially we reduce offline evaluation to the problem of answering the
question: ``What would an ideal system do?''
Once we determine the ideal ranking for a
query~---~or rather a set of equally ideal rankings~---we
apply a rank similarity measure to determine the compatibility
of an actual ranking generated by a ranker to this ideal.
As an offline evaluation measure,
compatibility is particularly suited to partial preferences,
since the weak ordering induced by partial preferences can be directly
interpreted as a set of ideal rankings.

In the current paper,
we extend our prior work to consider assessment methods for partial preferences.
Starting from a pool of items,
we examine  methods for narrowing this pool to the top-$k$ items,
identifying and ordering these items,
while minimizing the cost and effort required.
We compare two methods.
The first assumes dedicated and motivated assessors,
employing a tournament structure.
The second crowdsources preference judgments through Mechanical Turk.
For both methods, we start with an initial set of graded judgments as
a first step in narrowing the pool.

\begin{figure*}[t]
\begin{tabular}{c|l}
{\bf Grade} & {\bf Guidelines}\\
\hline
\hline
\begin{minipage}[c]{0.2\textwidth}
\centering
{\bf 4}\\
Fully meets
\end{minipage} &
\begin{minipage}[c]{0.75\textwidth}
\ \\
The passage is a perfect answer for the turn.
It includes all of the information needed to fully answer the turn in the
conversation context.
It focuses only on the subject and contains little extra information.\\
\end{minipage}\\
\hline
\begin{minipage}[c]{0.2\textwidth}
\centering
{\bf 3}\\
Highly meets
\end{minipage} &
\begin{minipage}[c]{0.75\textwidth}
\ \\
The passage answers the question and is focused on the turn.
It would be a satisfactory answer if Google Assistant or Alexa returned
this passage in response to the query.
It may contain limited extraneous information.\\
\end{minipage}\\
\hline
\begin{minipage}[c]{0.2\textwidth}
\centering
{\bf 2}\\
Moderately meets
\end{minipage} &
\begin{minipage}[c]{0.75\textwidth}
\ \\
The passage answers the turn, but is focused on other information that is unrelated to the question.
The passage may contain the answer, but users will need extra effort to pick
the correct portion.
The passage may be relevant, but it may only partially answer the turn,
missing a small aspect of the context.\\
\end{minipage}\\
\hline
\begin{minipage}[c]{0.2\textwidth}
\centering
{\bf 1}\\
Slightly meets
\end{minipage} &
\begin{minipage}[c]{0.75\textwidth}
\ \\
The passage includes some information about the turn,
but does not directly answer it.
Users will find some useful information in the passage that may lead to the
correct answer,
perhaps after additional rounds of conversation (better than nothing).\\
\end{minipage}\\
\hline
\begin{minipage}[c]{0.2\textwidth}
\centering
{\bf 0}\\
Fails to meet
\end{minipage} &
\begin{minipage}[c]{0.75\textwidth}
\ \\
The passage is not relevant to the question.
The passage is unrelated to the target query
\end{minipage}
\end{tabular}
\caption{
  Assessment guidelines for the TREC 2019 Conversational Assistance
  Task~\cite{dxc19}}.
\label{fig:guidelines}
\end{figure*}

We focus our effort on partial preferences for 
a question answering
task~---~the
TREC 2019 Conversational Assistance Track\footnote{\url{www.treccast.ai}}
(CAsT)~\cite{dxc19}.
For this task,
questions were collected into conversations of between 7 and 12 questions each.
Answers were drawn from a collection of passages derived from various Web
sources, including Wikipedia.
For each of the 479 test questions,
participating systems returned a ranked list of passages intended to answer
the question.
Submitted runs were pooled to a depth of 10, and 173 of the questions
were judged on the 5-point scale shown in Figure~\ref{fig:guidelines},
where a ``turn'' is a round of the conversation.
NDCG@3 formed the primary evaluation measure for the track.
Through the application of preference judging,
we aim to identify and order the top-five answers for these 174
previously judged questions.

\begin{figure*}[t]
\begin{minipage}[c]{0.9\textwidth}
{\bf MARCO\_2531173}:
Foods high in iron include:
1 red meat. 2 seafood. 3 organ meats, such as liver. 4 whole grains.
5 dried fruits. 6 nuts.  7 beans, especially lima beans.
8 dark green leafy vegetables, such as spinach and broccoli.
9 iron-fortified foods, such as breads and cereals (check the label) \\
\\
{\bf CAR\_9a6bd9a37b8e4643e1f1cb434b2d5fd40942e277}:
Broccolini is high in vitamin C (containing 100\% of daily intake)
and also contains vitamin A, calcium, Vitamin E, folate, Iron, and Potassium.
It has 35 calories per serving. \\
\\
{\bf MARCO\_1016779}:
Iron is a trace mineral that is important for healthy blood.
It helps red blood cells transport oxygen throughout your body,
and helps carry carbon dioxide out.
A deficiency in iron can lead to the condition known as anemia.
Food Sources: Red meat and egg yolks are high in iron.
\end{minipage}
\caption{
  Three of 55 passages assigned the top grade (``fully meets'')
  by the TREC 2019 Conversational Assistance Task for the question:
  {\em What foods contain high levels of iron?} (\#67.10).}
\label{fig:iron}
\end{figure*}

The questions from the TREC CAsT Track provide some excellent examples of the
problem that initially motivated us.
Figure~\ref{fig:iron} shows three passages that received the top 
grade (``fully meets'')
for the question {\em What foods contain high levels of iron?} (\#67.10).
When viewed in isolation,
any of these passages could reasonably be judged to answer the question,
but when placed side-by-side differences become clear.
While all three passages name foods that contain high levels of iron,
only the first passage provides a comprehensive list.
Figure~\ref{fig:cast} shows four passages that received the top grade
for the question {\em What is taught in sociology?} (\#79.1).
The first two passages provide direct answers,
while the third passage contains extraneous information
and the fourth is merely a disjointed list of topics.
The factors that make one ``fully meets'' passage better than another
are not captured in the ordinal relevance scale and assessing
guidelines shown in Figure~\ref{fig:guidelines}, but with preference
judgments, we can easily incorporate such factors into offline
evaluation and be more sensitive to systems able to
correctly rank the very best passages.  

\begin{figure}[t]
\begin{minipage}[c]{0.90\textwidth}
{\bf MARCO\_1568091}:
Sociology is the study of social life and the social causes and consequences
of human behavior.
In the words of C. Wright Mills,
sociology looks for the public issues that underlie private troubles.
Sociology differs from popular notions of human behavior in that it uses
systematic, scientific methods of investigation and questions many of the
common sense and taken-for-granted views of our social world...\\
\\
{\bf MARCO\_394140}:
What is Sociology?
Sociology is the study of human social relationships and institutions.
Sociology's subject matter is diverse,
ranging from crime to religion,
from the family to the state,
from the divisions of race and social class to the shared beliefs of a
common culture,
and from social stability to radical change in whole societies.\\
\\
{\bf CAR\_f62c5a5a0be476d8ba9ce5d956b519413d73eb71}:
Jennifer Conn used Snape's and Quidditch coach Madam Hooch's teaching methods
as examples of what to avoid and what to emulate in clinical teaching,
and Joyce Fields wrote that the books illustrate four of the five main topics
in a typical first-year sociology class:
``sociological concepts including culture, society, and socialisation;
stratification and social inequality; social institutions;
and social theory''.\\
\\
{\bf CAR\_5465fd5dd01cba27c7d792b6b6453ee3da101e03}:
sociology of aging - sociology of architecture - sociology of art -
sociology of the body - sociology of childhood - sociology of conflict -
sociology of deviance - sociology of development - sociology of disaster -
sociology of economic life - sociology of education - sociology of emotions -
sociology of the family -
%sociology of fatherhood - sociology of film -
%sociology of food - sociology of gender -
%sociology of giving - sociology of government -
%sociology of health and illness - sociology of the history of science -
%sociology of immigration - sociology of industrial relations -
%sociology of knowledge -
...
\end{minipage}
\caption{
  Four of the 14 passages assigned the top grade (``fully meets'')
  by the TREC 2019 Conversational Assistance Track for the question:
  {\em What is taught in sociology?} (\#79.1).
  The first passage contains two concatenated copies of this text;
  we show only one.
  The fourth passage has been truncated.}
\label{fig:cast}
\end{figure}

The remainder of the paper is organized as follows:
Section~\ref{sec:back} provides a review of prior work on preference judgments
for information retrieval evaluation.
Section~\ref{sec:partial} summarizes the foundational theory for compatibility,
which first appeared in our prior work~\cite{compat}.
Along with this prior work,
this section further explores the relationship between NDCG@$k$ and
compatibility when graded judgments alone are used to compute
compatibility.
Sections~\ref{sec:topk} to~\ref{sec:impact} represent the bulk of the new
research presented in this paper.
These sections define, validate, and test a crowdsourcing method for
preference judgments aimed at identifying the top-$k$ items.
We illustrate the impact of these top-$k$ preferences with runs from
TREC 2019 CAsT Track.

Code and preference judgments are available at
\url{https://github.com/claclark/compatibility}.
Details are provided as an appendix.
As part of institutional ethics review,
permission was given to include crowdsourced preference judgments in this
release without identifying information.

\section{Preference judgments}
\label{sec:back}

As far back as 1990,
\citet{rorvig90} argued for the superiority of preference judgments
as a tool for estimating document utility, as opposed to graded or binary
relevance judgments, explicitly recognizing that this utility may reflect
differences beyond that possible with ordinal relevance scales.
That paper raises the transitivity of preferences as a necessary requirement
for this utility estimation,
and it reports experiments demonstrating that document preference judgments
do exhibit the required transitivity.
Rorvig also outlines a procedure for constructing a test collection based on
preference judgments, while noting that this test collection
``would cost a great deal more to build than current collections,''
due to the large number of judgments required.
\citet{fs91}
also eschew absolute relevance in favor of preference judgments,
since human assessors are able to make relative comparisons more easily
and consistently.

In a 1995 paper,
\citet{yao95} proposed preferences judgments as a solution to the
difficulties already then encountered in attempts to define and interpret
ordinal relevance scales, which in some cases might suggest, for example,
``that a document with grade~2 is equivalent to two documents with grade one.''
Under Yao's proposal, preference judgments define a \emph{weak ordering} on the
collection, where items may be tied.
Just as we propose in this paper,
this weak ordering might be derived from direct pairwise comparisons or from
ordinal relevance grades,
avoiding the need to directly interpret grades as relevance values.
Effectiveness is then measured by computing the distance between this weak
ordering and a ranking generated by a search system.
Yao defines axioms required for this distance metric,
including the usual mathematical properties required of any distance metric.
Our compatibility measure, defined in Section~\ref{sec:partial},
follows this suggestion,
using rank similarity measures to compare ideal and system rankings.

More recently, Carterette and Bennett, along with various collaborators,
published a series of papers aiming to establish preference judgments
as a practical approach to offline search
evaluation~\cite{cb08,cbc08,cbcd08,zc10,cc12,cc13,cbch13}.
\citet{cbcd08}
provides evidence that preference judgments are generally transitive,
so that $O(n^2)$ judgments are not required for a pool of $n$ items.
They further recognize that prejudging non-relevant documents allows
these documents to be excluded from the pool for preference judging,
further reducing effort.
\citet{cbc08} describe the creation of one of the few test collections based
on preferences.
Along with \citet{cb08},
these papers propose evaluation measures based on the discordant pairs
in an actual ranking.

\citet{zc10} crowdsource preference judgments for search page layouts,
providing advice that informs our current effort.
\citet{cc12} employ preference judgments to generate an ideally diverse
ranking.
\citet{cc13} extend this work to define a evaluation measure for novelty
and diversity based on preference judgments.
\citet{cbch13} present an active learning approach to inferring a ranking
from crowdsourced preference judgments.

\citet{ra11} refer to the practice of inferring preferences from individual
relevance judgments~---~both to train rankers and for evaluation~---~as
the ``IR detour''.
Through experiments on human subjects they conclude that
``the validity of taking the IR detour is questionable.''
They propose an active learning method for reducing the number of preference
judgments.
In particular,
they propose focusing preference judgments on identifying the top-$k$ items,
although they do not explore this proposal in detail.
They also provide an overview of some of the earlier work in the large
body of literature related to preference judgments for learning-to-rank.
This literature includes research specifically focused on top-$k$
learning-to-rank methods~\cite{xll09,lngc13,sglc12}.

Another large body of literature explores methods for crowdsourcing
both graded and preference judgments~\cite{ars08,judges08,ly11}.
\citet{mmst17} crowdsource relevance magnitudes through a process in which
assessors view a series of documents and estimate relevance relative to
the previously seen document.
Their results call into question the standard practice of converting
relevance grades into gain values for the purpose of computing NDCG.
\citet{hb17a,hb17b} explore the transitivity of crowdsourced preference
judgments and propose an algorithm based on a randomized quicksort to reduce
judging effort by allowing ties.
\citet{ymt18} compare preference, absolute and ratio judgments through
a large crowdsourced experiment,
concluding that crowdsourced preferences provided similar outcomes as
dedicated assessments when comparing rankers.

\citet{bawgpa13} propose methods for converting preference judgments to 
relevance scores by adapting the ELO ratings used for chess and other games.
\citet{kkz13} provide evidence that preference judgments can capture 
differences beyond traditional topical relevance,
such as authority and recency.
\citet{az14} employ a classifier to reduce the effort associated with
preference judgments.
\citet{kuhl19} explore interaction methods for collecting preference
judgments.
\citet{krg18} augment star ratings with preference judgments in a recommender
system.

In a recent SIGIR 2020 paper, \citet{sz20} propose and explore two broad
families of measures intended to support preference judgments.
The first family is based on counts of concordant pairs,
generalizing and extending ideas proposed by \citet{cbcd08} and \citet{cb08}.
The second family converts preference judgments to gain values for use
with traditional graded measures.
A unique aspect of these measures is that they work directly from a
collection of preference judgments,
and do not require assumptions of transitivity.
As part of this work,
the authors released an exhaustive set of preference judgments for an
NTCIR task.
Overall,
their work demonstrates several important advantages of preference judgments,
especially their closer agreement with SERP preferences,
but questions remain regarding the costs and sensitivity of measures based
on preference judgments.

In another recent SIGIR 2020 paper, 
\citet{xie20} apply preference judgments to image search.
They recognize that,
like may other domains,
judging images on absolute scales of relevance poses difficulties due to the
``multi-dimensional nature of relevance for images''.
Most notably, they extend the measures of~\citet{cbcd08} to accommodate
the grid-based result presentations common in image search.
Applying compatibility to image grids would requires the definition of 
similarity measure suitable for comparing
grids~---~an interesting problem in-and-of itself.
However, they conclude with the familiar concern that preference-based
evaluation may
``require a larger number of judgments than relevance-based evaluation even
after assuming transitivity,''
and wonder rhetorically
``How to reduce the number of judgments without affecting the effectiveness
of the preference-based metric?''

Given the quality and breadth of this prior research,
it is perhaps surprising that preference judgments are not yet standard for
offline search evaluation.
Many of the key ideas we employ in this paper have been explored,
or at least proposed, in this prior work.
We view the primary contribution of this paper and our related
papers~\cite{compat, beyond} as consolidating and simplifying this prior work
to establish the practical utility of preference judgments.
In particular,
we focus preference judgments on the top items to maximize impact while
minimizing judging effort.
In addition,
we further establish maximum similarity to an ideal ranking as a simplified
framework for offline evaluation,
accommodating traditional graded, judgements, preference judgments,
and factors beyond those of typical ordinal relevance scales.  

\section{Computing compatibility}
\label{sec:partial}

\subsection{Compatibility}
\label{sec:compat}

Computing compatibility requires two choices:
1) a choice of rank similarity measure to compare rankings,
and 2) a definition of an ideal ranking, 
which might be a single ranking or a set of equally ideal rankings.
For rank similarity we use RBO
because its properties make it ideally suited for comparing rankings
(see Section~\ref{sec:rbo}).
For the experiments in this paper,
we define the ideal rankings for a query by a set of equivalence classes,
or ''effectiveness levels'',
where each effectiveness level contains one or more items.

Let $\{L_1, L_2, ..., L_T\}$
be the set of effectiveness levels for a query.
The effectiveness levels are ordered so that $L_1 < L_2 < ... < L_T$,
with $L_T$ being the top level.
Unlike traditional graded assessment,
the number of levels $T$ can vary from query to query.
We define an extra level $L_0$ containing all items not appearing in
another level.
We define an ideal ranking as any ranking containing all the items in $L_T$,
in any order,
followed by all the items in $L_{T-1}$,
in any order,
and so on down to $L_1$.
The items in $L_0$ are not included.

If we have graded judgments,
these effectiveness levels correspond exactly to the grades,
with $L_0$ containing items that are non-relevant, spammy, unjudged, etc.
If we have an ideal ranking exactly defined by a top-$k$ ranking of items,
then we have $T=k$, with the first item alone in $L_k$,
the second item alone in $L_{k-1}$, etc.
We can also combine a top-$k$ ranking with graded judgments by ordering
the top-$k$ items first and ordering the remaining items in the grades
below them.
In this paper, we do all three.

Together, a set of equivalence levels defines a set of ideal rankings
containing $|L_T|! \times |L_{T-1}|! \times ... \times |L_1|!$ elements.
If equivalence levels are based on grades,
the size of this set can be a million or more for a typical TREC task.
For TREC 2019 CAsT questions, the size of this set ranges from 192 ideal
rankings up to 26,842,725 ideal rankings,
with an average above two million.
In contrast, with a top-$k$ ranking,
the sole element in the set can precisely specify what the searcher should see.

Fortunately, regardless of the number of ideal rankings,
we do not need to generate all of them to determine the ideal ranking.
This maximum will be obtained by the ideal ranking that has all the items in
each level ordered according to the actual ranking,
maximizing the number of concordant pairs~\cite{compat,beyond}.
For items not appearing in the actual ranking,
they should be placed last in the level in any order.
Once we have chosen a rank similarity measure and
defined a set of ideal rankings,
we compute compatibility as the maximum similarity between members of the
set and the actual ranking generated by a ranker we wish to measure.

\subsection{Rank biased overlap}
\label{sec:rbo}

While in principle any rank similarity measure could be used to compute
compatibility, we employ Rank Biased Overlap (RBO).
By design, its properties make it ideally suited for this purpose.
In creating RBO,
\citet{wmz10} carefully identified and specified the requirements of rank
similarity for what they call indefinite rankings,
such as the output of rankers.
For example,
when comparing an actual ranking generated by a ranker to an ideal ranking,
the top ranks matter more and should be given greater weight.
The ideal ranking may be relatively short~---~just
the top-5, for example~---~while
the actual ranking may be much longer~---~up
to 1000 passages for TREC CAsT experimental runs because all the items
appearing in the ideal ranking may not appear in the actual ranking.
RBO allows us to meaningfully compare rankings with differing length and
content.
While we could certainly employ or invent other rank similarity measures,
they would still need to satisfy the requirements of ~\citet{wmz10}.
Further discussion can be found in our related paper~\cite{compat}.

Using RBO, we compute compatibility between an ideal ranking $I$ and an
actual ranking $R$ as follows:
Let $I_{1:i}$ denote the top $i$ items in $I$,
and let $R_{1:i}$ denote the top $i$ items in $R$.
We define the {\em overlap} between $I$ and $R$ at depth $i$ as the
size of the intersection between these lists at depth $i$:
$\left|I_{1:i} \cap R_{1:i}\right|$.
We define the {\em agreement} between $I$ and $R$ at depth $i$ as the overlap
divided by $i$.
RBO is then a weighted average of the agreement across depths from
1 to $\infty$, as follows:
\begin{equation}
\label{eqn:RBO}
\mbox{RBO}(R, I) =
  (1 - p)
    \sum_{i=1}^{\infty}
      p^{i - 1} \frac{\left|I_{1:i} \cap R_{1:i}\right|}{i}.
\end{equation}
The parameter $0 < p < 1$ represents searcher patience or persistence,
with larger values representing more persistent searching.
For practical purposes,
the summation is computed down to sufficient depth
so that $p^{i - 1}$ is close to zero
and we reach the bottom of both the ideal and actual rankings.
We go down to depth 1000 for this paper.
Please see ~\citet{wmz10} for further discussion.

\begin{table}[t]
\begin{tabular}[t]{rc|cc|c}
\multicolumn{2}{c|}{NDCG} & \multicolumn{2}{c|}{Compatibility}\\
$k$  & sensitivity &  $p$ & sensitivity & Kendall's $\tau$\\
\hline
 3   &    71.7\%   & 0.80 &    71.0\%   &      0.907      \\
 5   &    72.6\%   & 0.85 &    73.5\%   &      0.920      \\
10   &    76.6\%   & 0.90 &    76.5\%   &      0.910      \\
20   &    78.2\%   & 0.95 &    79.4\%   &      0.956
\end{tabular}
\caption{
  Sensitivity and consistency of NDCG and compatibility on 
  TREC 2019 Conversational Assistance Track automatic runs
  when ideal rankings are based on graded judgments only.
}
\vspace*{-\baselineskip}
\label{tab:scMore}
\end{table}

\subsection{Consistency and sensitivity}

Along with other analyses,
we compare evaluation measures in terms of their \emph{consistency}
and \emph{sensitivity}.
By consistency we mean the degree to which evaluation measures recognize
the same differences between rankers.
By sensitivity (often called ``discriminative power'') we mean the ability of
evaluation measures to recognize significant difference between
rankers\cite{sakai06}.

We measure consistency using Kendall's $\tau$,
comparing the ordering of runs under two measures.
Kendall's $\tau$
has long been employed to measure consistency in information retrieval
evaluation~\cite{voor98}.
We measure sensitivity following the approach of \citet{sakai06}
but using paired t-tests rather than bootstraps,
following the approach of \citet{ymt18}.
We take all pairs of experimental runs and compute a paired t-test between
them under each measure.
A pair with $p < 0.05$ is considered to be \emph{distinguished}.
Sensitivity is then:
\begin{equation}
\mbox{sensitivity} =
  \frac{\mbox{\# of distinguished pairs}}{\mbox{total pairs}}
\end{equation}
Please note that sensitivity reflects a property of the evaluation measure and
that~---~because there is no Bonferroni or other correction~---~some
of the distinguished pairs may not represent actual significant differences.
Sensitivity is really a measure of ``best case'' performance of the
evaluation measure,
allowing us to compare one measure to another.

\begin{figure}[t]
\centering
\includegraphics[width=2.6in, keepaspectratio]{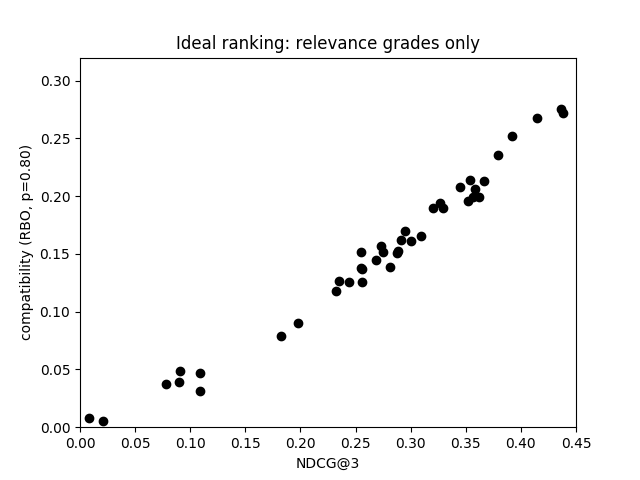}
\includegraphics[width=2.6in, keepaspectratio]{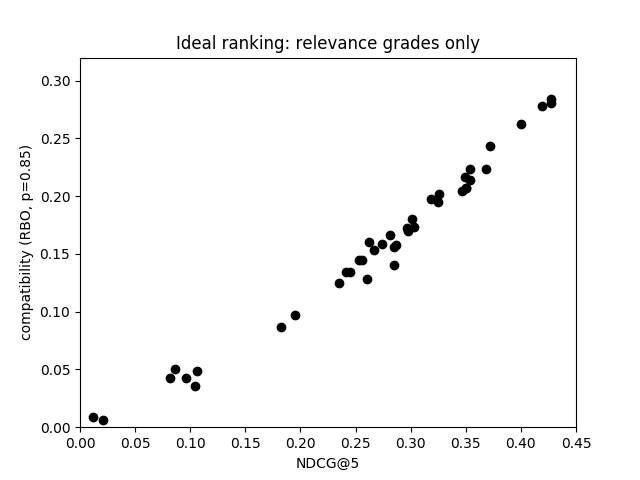}
\includegraphics[width=2.6in, keepaspectratio]{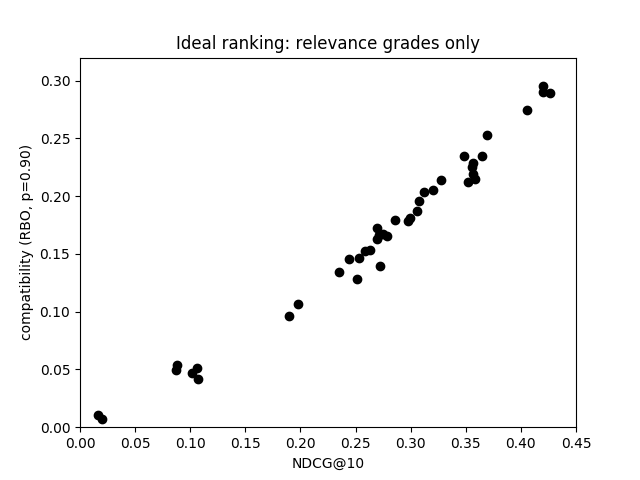}
\includegraphics[width=2.6in, keepaspectratio]{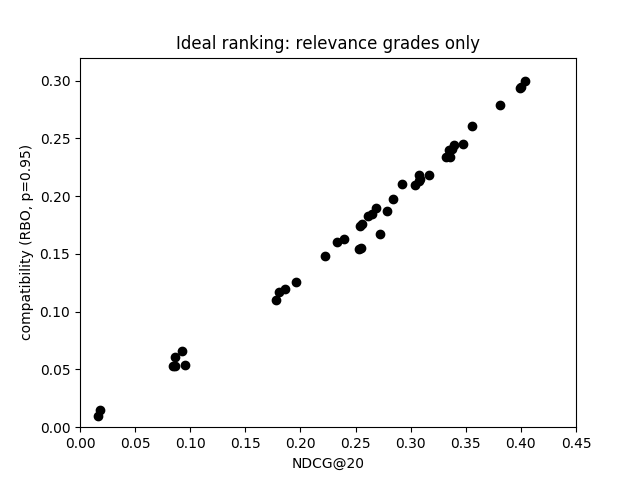}
\caption{
  The relationship between NDCG and compatibility on 
  TREC 2019 Conversational Assistance Track automatic runs
  when ideal rankings are based on graded judgments only.
  The relationship between the $k$ and $p$ parameters are discussed in the text.
  Even though compatibility does not convert grades to gain values,
  the relationship appears nearly linear, with Kendall's {\bf $\tau > 0.9$}
  in all cases.}
\label{fig:graded}
\end{figure}

\subsection{Compatibility with grades only}

As detailed in Section~\ref{sec:compat} grades alone can be used
to define a set of ideal rankings,
allowing compatibility to be computed.
For the TREC 2019 CAsT task,
there are four effectiveness levels.
The top effectiveness level $L_4$ contains all passages judged ``fully meets'',
$L_3$ contains all passages judged ``highly meets'',
$L_2$ contains the ``moderately meets'' passages,
and $L_1$ contains the ``slightly meets'' passages.
Figure~\ref{fig:graded} compares compatibility and NDCG on the
42~automatic runs from TREC 2019 CAsT.
While NDCG@3 was primary evaluation measure reported for
TREC 2019 CAsT~\cite{dxc19},
we report NDCG@$k$ for values of $k \in \{3, 5, 10, 20\}$.
The relationship between these measures appears nearly linear.
Values for Kendall's $\tau$ are all greater than $0.9$
(see Table~\ref{tab:scMore})
with few inversions in the higher scoring runs.

For these values of $k$,
we tuned the value of $p$ to provide the best match, in terms of consistency,
between RBO-based compatibility and NDCG@$k$.
Since larger values of $p$ give larger weight to items deeper in the ranking,
we might expect larger values of $p$ to correspond to larger values of $k$.
We tuned $p$ on the four years of test collections from the TREC Web Track,
which we employed in our prior work~\cite{compat}.
Tuning was entirely manual;
we tried four or five values for each $k$ before settling on values of 
$p$ that provided good consistency in terms of Kendall's $\tau$
across the four years.
We did not tune on the TREC CAsT Track data used in the current paper.
These values for $p$ for compatibility provide roughly the same
sensitivity as the corresponding values of $k$ for NDCG.
Table~\ref{tab:scMore} shows the correspondence between NDCG and RBO-based
compatibility when ideal rankings are based only on grades.

\section{Identifying the top-$k$}
\label{sec:topk}

Our goal is to identify the top-$k$ items for each query while minimizing
effort.
We follow a multi-step approach,
depending on if the assessment will be completed by dedicated assessors or
by crowdsourced assessors.
We assume that dedicated assessors will be more focused and reliable than
crowdsourced assessors,
so we build more redundancy into the crowdsourced process.
Our overall approach is to favor simplicity.
It can be summarized as follows:
\begin{enumerate}
\item
Perform an initial graded assessment pass to ``thin the herd'',
producing a reduced candidate pool ${\mathcal{C}}$,
with $|\mathcal{C}| \geq k$
to focus preference judgments on the most promising items
(Section~\ref{sec:herd}).
\item
If dedicated assessors are to be used,
we structure assessment as a single-elimination tournament
(Section~\ref{sec:me}).
\item
If crowdsourced assessors are to be used we follow a two-stage process,
with the first stage reducing the size of the candidate pool and the
second stage determining the final order
(Section~\ref{sec:them}):
\begin{enumerate}
\item
While the size of the candidate pool is greater than some threshold $F$,
where $|C| > F > k$,
we generate random pairings of candidates,
so that each candidate is paired with $P$ or $P+1$ other candidates,
where $F > P > k$.
These pairings are then judged by crowdworkers,
for some threshold $P > k$.
Items losing more than a majority of pairings are eliminated, and we repeat.
\item
Once the size of the candidate pool is less than or equal to $F$,
we pair all remaining candidates with all other remaining candidates,
which are judged by crowdworkers.
Items are then ranked by the number of pairs they win,
and we cut to the top $k$.
In the case of ties at rank $k$,
we keep all candidates with the tied score,
so that in some cases the size of the final ideal ranking will be larger than
$k$.
\end{enumerate}
\end{enumerate}
For the experiments in this paper, we use $k = 5$, $F = 9$, and $P = 7$.
The values for $F$ and $P$ were based on a pilot test,
intended to keep our costs under \$4,000.

\subsection{Thinning the herd}
\label{sec:herd}

We start with an initial graded assessment,
giving us an initial candidate pool of higher quality items
and avoiding unnecessary preference judgments against lower quality items,
particularly non-relevant items.
These initial judgments could be crowdsourced or generated by dedicated
assessors.
If we assume $g$ grades,
with $\mathcal{G}_0$, $\mathcal{G}_1$,...  $\mathcal{G}_g$
as the sets of items for each grade,
we compute $\mathcal{C}$ as follows: \begin{minipage}[t]{3.5in}
\begin{tabbing}
\ \ \ \ \ \ \ \ \= \ \ \ \= \ \ \ \= \ \ \ \=\\
\> $i \leftarrow g$\\
\> $\mathcal{C} \leftarrow \emptyset$\\
\>while $|\mathcal{C}| < k$ and $i > 0$:\\
\>\>$\mathcal{C} \leftarrow \mathcal{G}_i \cup \mathcal{C}$\\
\>\>$i \leftarrow $i - 1
\end{tabbing}
\end{minipage}

\begin{figure}[t]
\centering
\includegraphics[width=3.3in, keepaspectratio]{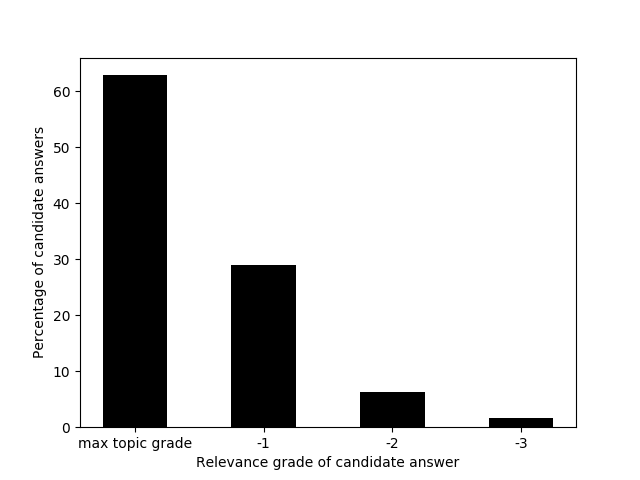}
\caption{
  Grades of passages selected for the candidate pool relative
  to the top grade for the question.
}
\label{fig:candidates}
\end{figure}

For the TREC 2019 CAsT task,
experimental runs were pooled down to depth~10 for assessment.
A total of 29,350 passages were judged on a 5-point scale,
from ``fully meets''(4) down to ``fails to meet''(0),
Of these, 8,120 passages were assigned a positive grade.
Running the algorithm above on the passages with a positive grade
gives an initial candidate pool of 2,673 passages.
The number of candidates vary by question up to a high of 112 for question
\#67.8.
Of the 173 questions, 57 had an initial candidate pool with
$|\mathcal{C}| \leq F$,
so that for crowdsourced assessments,
these candidates immediately moved to the second stage.
As shown in Figure~\ref{fig:candidates},
not all candidates came from the top grade for that question.
More than a third came from below the top grade,
with a just over 1\% coming from three levels lower.
Since we are depending on the grades to build the initial candidate
pool, it is certainly possible that some of the top answers were missed by this
process; we further discuss this possibility later in the paper.

\begin{table}[t]
\begin{tabular}{cc|rrr}
Topics & $k$ & official & extra & \% extra\\
\hline
173 &  3 & 29,350 &  3,456 & +11.78\% \\
  " &  5 &      " &  5,429 & +18.50\% \\
  " & 10 &      " & 10,691 & +36.43\% \\
\end{tabular}
\caption{
  Upper bound estimates of extra judging effort to identify top-$k$ items for
  the TREC 2019 CAsT task with dedicated and reliable assessment.}
\label{tab:extra}
\end{table}

\subsection{Dedicated assessment}
\label{sec:me}

If we have reliable and dedicated assessment,
undertaken by a relatively small number of individuals who understand the task,
we can use a single-elimination tournament structure, or heap,
to determine the top-$k$ items with no more than
$|\mathcal{C}| + (k - 1)\lceil\log(|\mathcal{C}|)\rceil$ preference judgments
(not a tight bound).
Using this formula, Table~\ref{tab:extra} provides an estimate of the
preference judgments required for TREC 2019 CAsT for various values of $k$.

To provide a basis for comparison with crowdsourcing results,
the authors applied this approach to
identify a single top answer for each of the questions.
Over the course of several weeks, and requiring nearly 40 hours,
we completed the first round of the single-elimination tournament.
In total we made 3,743 preference judgments.
This total includes some judgments completed for pilot tests,
consistency checks, interface refinement, and similar activities.
Without these extra judgments,
only 2,498 judgments would be required to fully judge this pool,
requiring approximately 27 hours of effort.
This process gave us a top answer for each question.
In the next section,
we use the top answers produced by this dedicated assessment process
to help validate the crowdsource assessment.

\subsection{Crowdsourced assessment}
\label{sec:them}

As described above, crowdsourcing proceeds in two stages:
a) a pool reduction stage,
intended to reduce the size of the candidate pool below some threshold $F$,
after which we
b) compare all remaining candidates with each other,
ranking the candidates according to the number of pairings in which they win
and cutting to the top $k$.
During the pool reduction stage each candidate is randomly paired with
$P$ or $P+1$ other candidates,
with no repeated pairing.
We use a brute-force algorithm to generate random graphs for this purpose.
Candidates failing to win a majority of pairings are culled.
If the size of the pool is still greater than $F$,
we repeat the process.
On the TREC 2019 CAsT candidate pool,
each iteration of this process reduced the size of the pool by roughly half.

During the second stage all candidates are paired against each other,
giving up to $F(F-1)/2$ pairs.
By fully judging all pairs, we hope to improve the consistency of the top-5.
However, if these second-stage judgments are not fully transitive,
ties can result.
If the ties occur at rank $k$, we include all items tied at that rank.
Otherwise, we cut to the top $k$.
Ties also mean that some effectiveness levels will contain multiple items.

For the TREC 2019 CAsT passages,
we used Amazon's Mechanical Turk to recruit and pay crowdsourced workers.
Workers were required to live in the U.S.
and to have completed at least 1,000 HITs with an approval rating above 95\%.
Preference judgments were grouped into sets of 10,
forming a single HIT for which we paid \$2.00 to the worker,
as well as a fee of \$0.40 to Amazon.
Each HIT also included three challenge questions,
pairing a random passage from the candidate pool against a random
non-relevant passage.
HITs by workers failing a challenge question were discarded;
these workers were paid but excluded from further work.

In total,
crowdsourcing cost \$3,879.60 for 15,349 preference judgments,
including some pilot judgments and HITs excluded by the challenge questions.
This corresponds to an average cost of just over \$0.25 per preference judgment.
Overall, preference judging required 52.3\% additional judgments beyond the
29,350 initial graded judgments.
Assuming the same average cost for a graded judgment and a preference judgment
gives us a cost estimate of under \$12K for the full assessment exercise.

\begin{figure}[t]
\centering
\includegraphics[width=0.95\textwidth, keepaspectratio]{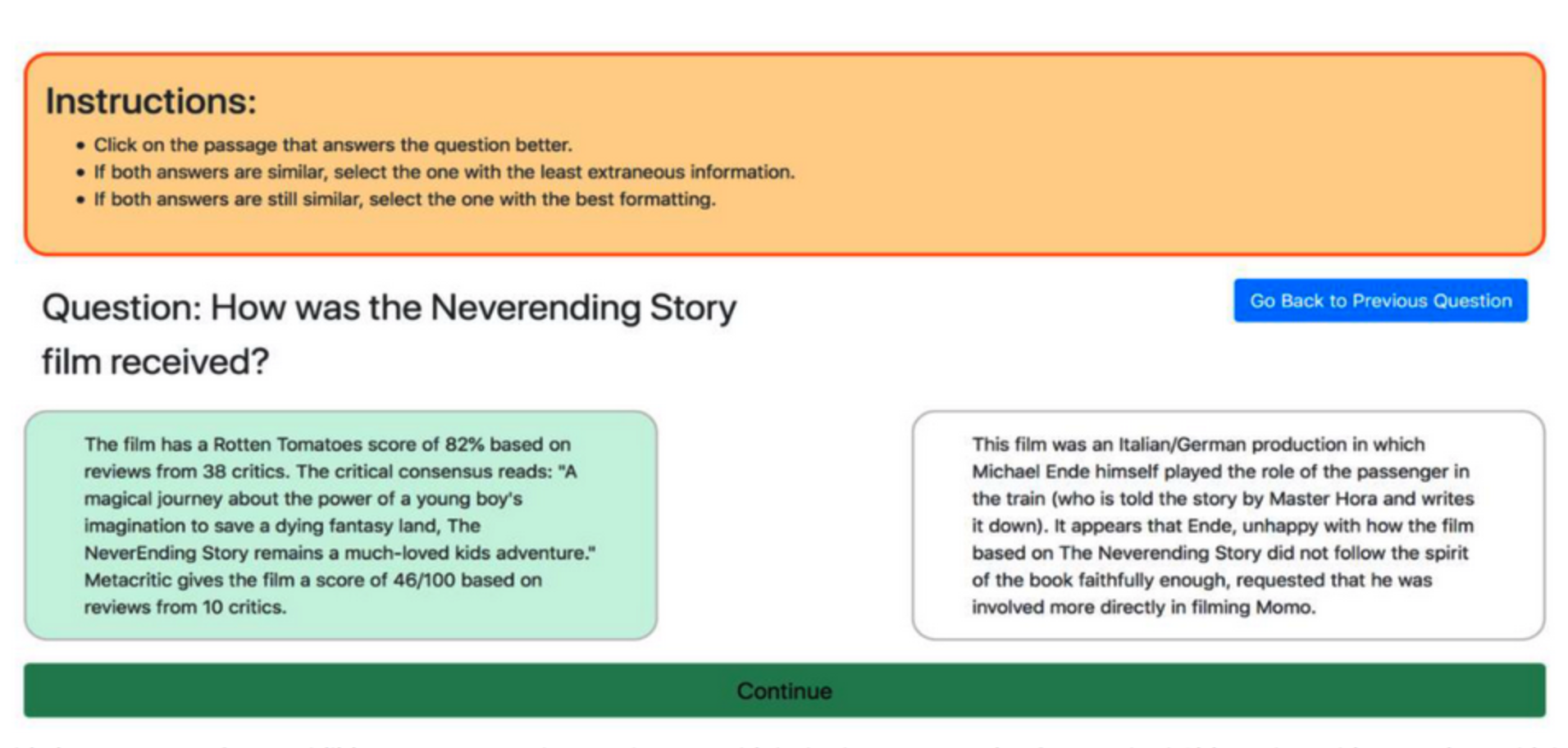}
\caption{
  Example assessment task for question \#33.3 from the
  TREC 2019 Conversational Assistance Track.}
\label{fig:interface}
\end{figure}

Figure~\ref{fig:interface} provides an example of the judging interface.
As was done for the official assessments,
our assessments used the manually re-written questions supplied by the track,
rather than the raw utterances from the conversations.
Unlike the track assessment, questions were shown in isolation,
rather than conversation order, a possible confound.

We kept the instructions simple,
asking workers to identify the passage that ``best answers the question.''
To break ties,
we asked them to choose the one with the least extraneous information.
All else being equal,
we asked them to choose the one with the ``best formatting'',
a phrasing we hoped would encourage them to choose on the basis of any
passage-specific factors we they believed to be important.
We deliberately did not allow assessors to indicate ties.
As much as possible,
we encouraged workers to indicate a preference,
with the goal of making distinctions between the top answers.
The simplicity and conciseness of these instructions can be compared with
the assessment guidelines required for graded
assessment in Figure~\ref{fig:guidelines}.

This study was approved by our institutional review board,
who also approved the release of the preference judgments without
personally identifying information.
As required by our institution,
the payment of \$2.00/HIT was intended to provide compensation equal to or
greater than minimum wage.
Based on our dedicated assessment experience,
we estimated a rate of one judgment per minute or higher,
or roughly one HIT every 10~minutes.
This rate translates to an estimated payment of \$12.00 for an hour's work,
consistent with our local minimum wage.

\section{Assessment Comparison}

\begin{figure}[t]
\centering
\includegraphics[width=3.3in, keepaspectratio]{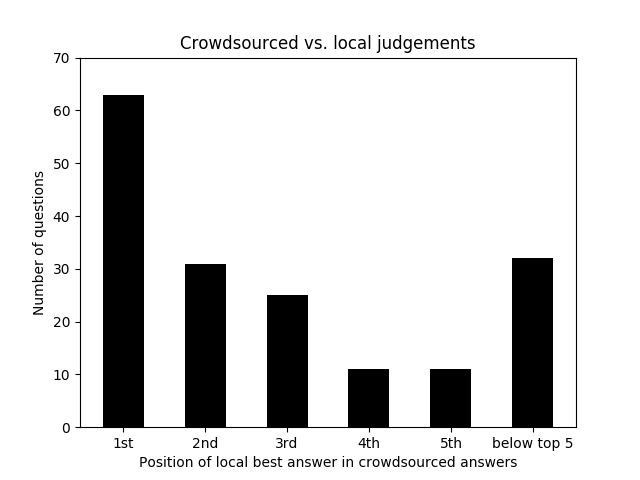}
\caption{
    Comparison between local and crowdsourced judgments.}
\label{fig:compare}
\end{figure}

Having completed both a crowdsourced assessment for the top-5 answers
and a dedicated assessment for the top answer
(which we call the ``local answer'' for short)
we can compare the two approaches.
Figure~\ref{fig:compare} shows the result.
For 63 questions (36\%) the two assessment methods produced the same top answer.
For example, both assessment methods identified the first passage in
Figure~\ref{fig:iron} as the top answer.
For 141 questions (82\%) the local answer from the dedicated assessment
appeared in the top-5 from the crowdsourced assessment.
For example,
of the passages in Figure~\ref{fig:cast} the first passage was selected by
crowdworkers as the top answer.
The second passage was ranked second by the crowdworkers,
but was the top local answer.
For 32 questions the local answer did not appear in the top
five crowdsourced answers at all.
In general, the crowdworkers appeared to prefer more direct answers,
and appeared less tolerant of longer passages than the dedicated assessors.

\begin{figure}[t]
\centering
\includegraphics[width=3.3in, keepaspectratio]{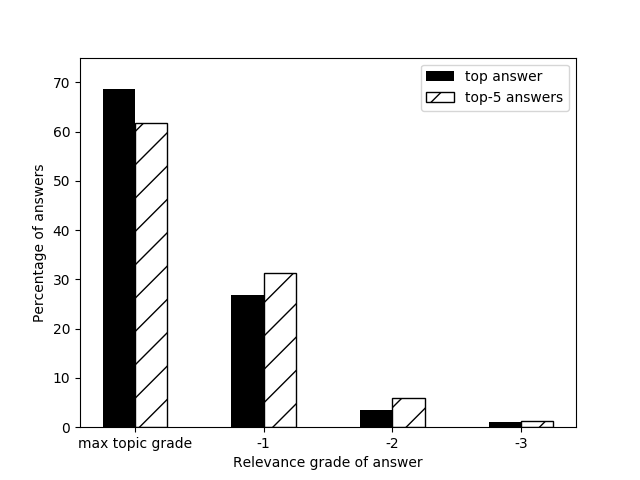}
\caption{
  Grades of top crowdsourced answer relative to the
  top grade for the question.
}
\label{fig:top}
\end{figure}

Figure~\ref{fig:top} compares the crowdsourced assessments with the original
graded assessments.
Over 68\% of the top-1 crowdsourced answers came from the highest
grade for the question,
which varied from question to question.
Over 61\% of the top-5 crowdsourced answers came from the highest
grade.
Nonetheless, the remaining answers came from lower grades.
Since we only added passages from lower grades when they were
needed to grow the candidate pool to sufficient size,
this outcome suggests that our initial strategy for ``thinning the herd''
may have missed some answers that the crowdworkers would have placed in the
top 5.

The values for $F$ and $P$ were chosen to keep us within an assessment budget
of \$4,000.
After running a pilot study with 10\% of the questions picked at random,
we set $F = 9$ and $P = 7$,
which kept us under budget.
Nonetheless, even if we assume fully consistent crowdworkers,
there is a small chance that some of the top-5 items might be missed.
The worst case occurs with a candidate pool $|\mathcal{C}| = F + 1 = 10$.
In this case with $P = 7$ there is more than a 12\% chance that the
fifth-best answer will be paired with all the top-4 answers
and would fail to win a majority of its pairings.
However, once the size of the candidate poll $|C| \leq F$,
and we have moved to the second stage,
all pairs are assessed, providing redundancy for the final top-5 ordering.

Overall, the assessment methods produced consistent,
but not identical, results.
By basing an initial pass on the original graded judgments,
we may have missed answers that crowdworkers would have placed in the top 5.
Larger value of $F$ and $P$ may have produced more consistent results,
although at greater cost.
However, assuming that the top-5 crowdsourced answers provide
an acceptable approximation to the true top-5,
we can move on to examine the impact of partial preferences
on runs submitted to the TREC 2019 CAsT Track.

\section{Impact of Partial Preferences}
\label{sec:impact}

\begin{figure}[t]
\centering
\begin{minipage}[c]{2.9in}
\includegraphics[width=2.9in, keepaspectratio]{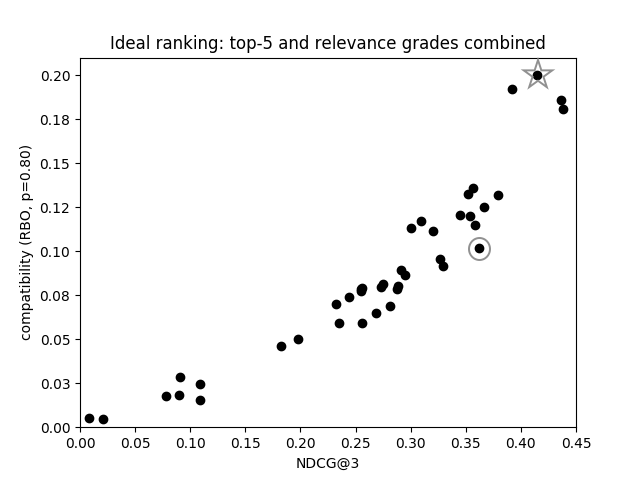}
\end{minipage}
\begin{minipage}[c]{2.5in}
\begin{minipage}[t]{2.5in}
\includegraphics[width=2.5in, keepaspectratio]{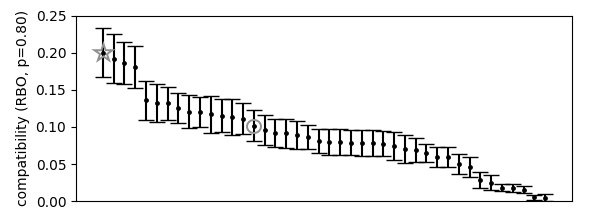}
\end{minipage}
\begin{minipage}[b]{2.5in}
\includegraphics[width=2.5in, keepaspectratio]{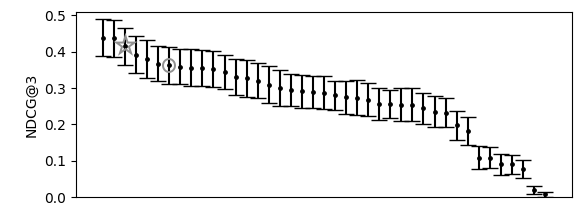}
\end{minipage}
\end{minipage}
\begin{minipage}[c]{2.9in}
\includegraphics[width=2.9in, keepaspectratio]{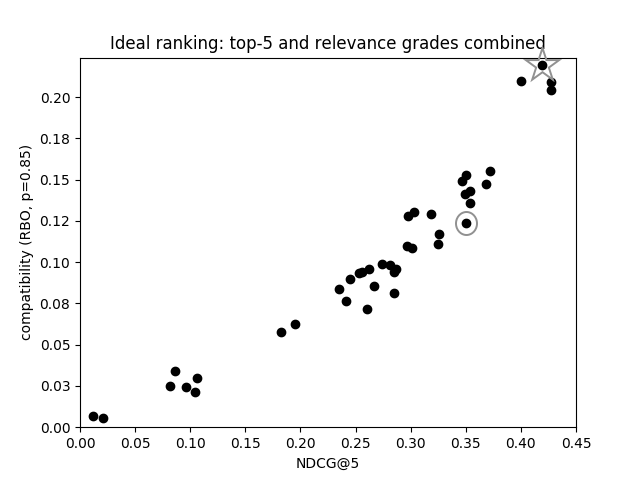}
\end{minipage}
\begin{minipage}[c]{2.5in}
\begin{minipage}[t]{2.5in}
\includegraphics[width=2.5in, keepaspectratio]{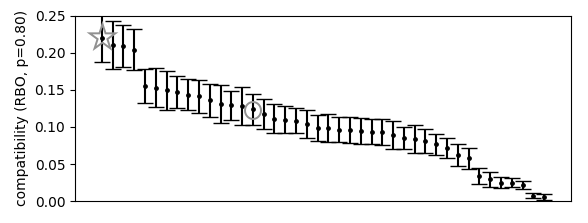}
\end{minipage}
\begin{minipage}[b]{2.51in}
\includegraphics[width=2.5in, keepaspectratio]{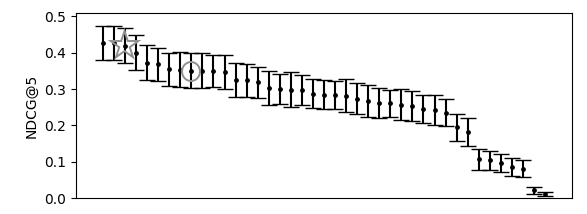}
\end{minipage}
\end{minipage}
\caption{
  The relationship between NDCG and compatibility on 
  TREC 2019 Conversational Assistance Track runs
  when ideal rankings are based on a combination of
  crowdsourced top-5 answers and the original graded judgments.
  The plots on the right sort runs by score under different measures and 
  show 95\% confidence intervals.
  The top-four runs show significant differences not captured by
  grades alone.
  The runs marked with a star and a circle are discussed in
  Section~\ref{sec:impact}.
  }
\label{fig:cqx}
\end{figure}

The plots in Figure~\ref{fig:cqx} compares the performance of automatic runs
submitted to the TREC 2019 CAsT Track under
compatibility vs.\ NDCG.
For this comparison,
we create an ideal ranking by combining the crowdsourced top-5 answers
with the original graded judgments.
The top-five answers fill equivalence levels $L_9$ down to $L_5$;
graded judgments fill equivalence levels $L_4$ down to $L_1$.
This approach precisely specifies the top ranks,
the ones most likely to be seen by the searcher,
while still taking advantage of the grades to compare rankers.

The figure reports compatibility ($p = 0.80$) vs.\ NDCG@3 and 
compatibility ($p = 0.85$) vs.\ NDCG@5,
which were official measures reported at TREC 2019.
The four plots on the right of the figure show 95\% confidence intervals
for the four measures.
Under compatibility,
we see a clear separation between the top-4 runs and the remaining runs,
which is not captured by grades alone.
As shown in Table~\ref{tab:sc} the sensitivity of compatibility using this
ideal ranking is 76.5\% with $p = 0.80$ and 77.8\% with $p= 0.85$,
indicating that we are better able to recognize differences between rankers.

Compatibility provides insights not provided by NDCG.
The top four runs (by either measure) represent the most successful
of the numerous attempts by participants to apply BERT~\cite{bert} for
re-ranking answers.
Under compatibility we see a clear separation between these four top runs
and the other runs, which is not evident under NDCG.
The starred run (pgbert) produces the best score under
compatibility and third-best score under NDCG.
In addition to BERT for re-ranking,
it applied a transfer learning approach for question re-writing~\cite{dxc19}.
Of the other three runs in the top four,
one (pg2bert) is variant of the pgbert run from the same group.
The other two (h2oloo\_RUN2 and CFDA\_CLIP\_RUN7)
both apply doc2query for expansion,
as well as BERT for re-ranking~\cite{cfda19}.
Most of the remaining runs in the top ten also attempted to apply BERT for
re-ranking,
but under NDCG the distinction between these runs and the four top runs 
is less evident.
Digging into the details of the differences between the four top runs
and the remaining BERT-based runs suggests that the query expansion and
re-writing methods used for the four top runs may have been the crucial
factor in their relative success.

The circled run (clacBase) was the sole run in top ten to use only
traditional IR methods~\cite{clac19}.
In particular, it was the only run in the top ten not to re-rank with BERT.
Under NDCG@3, the starred run outperforms the circled run by +15\%,
which is not significant under a paired t-test ($p = 0.11$),
even before Bonferroni or similar correction.
Under the corresponding compatibility measure,
the starred run outperforms the circled run by +97\%,
with a p-value $< 10^{-6}$,
which remains significant even after the conservative Bonferroni correction.

Under NDCG,
we might conclude that the modern NLP methods used for the starred run
were providing only a modest and non-significant improvement over the
traditional methods.
Under compatibility,
with an ideal ranking that precisely specifies the preferred answers,
we see the more dramatic improvements we might expect from these modern methods.
The remainder of the top-ten runs,
plus several other runs that also apply BERT,
move ahead of this traditional run under compatibility,
which drops from 7th to 15th place under NDCG@3.

The circled run forms the baseline for runs by the same group that applied
BERT for re-ranking (clacBaseRerank).
Under NDCG@3, the baseline slightly outperforms the re-ranked run
(0.360 vs.\ 0.343) and this difference is not significant.
Under the corresponding compatibility measure the re-ranked run
significantly outperforms its baseline by over 18\% (0.102 vs.\ 0.121),
with a p-value of $< 0.03$.
The groups submitting these runs re-ranked only the top-32 passages from the
baseline run, 
which was not deep enough to provide a positive impact under NDCG,
but was sufficient to provide a positive impact under compatibility.

\begin{table}[t]
\begin{tabular}[t]{rr|cc|r}
Measure       & Judgments    & Sensitivity & Kendall's $\tau$ & \\
\hline
NDCG@3        & graded only  & 71.7\%      & -      &  \multicolumn{1}{c}{-}\\
compatibility ($p = 0.80$)
              & graded only  & 71.0\%      & 0.907  & Fig.~\ref{fig:graded}\\
"\ \ \ \      & combined     & 76.5\%      & 0.851  & Fig.~\ref{fig:cqx}   \\
"\ \ \ \      & top-5 only   & 73.3\%      & 0.814  & Fig.~\ref{fig:cq}    \\
"\ \ \ \      & best only    & 55.2\%      & 0.775  & Fig.~\ref{fig:cq1}   \\
\hline
NDCG@5        & graded only  & 72.6\%      & -      &  \multicolumn{1}{c}{-}\\
compatibility ($p = 0.85$)
              & graded only  & 73.5\%      & 0.920  & Fig.~\ref{fig:graded}\\
"\ \ \ \      & combined     & 77.8\%      & 0.851   & Fig.~\ref{fig:cqx}   \\
\end{tabular}
\caption{
  Sensitivity and consistency of NDCG and compatibility for various judgment
  sets examined in this paper.
  Kendall's $\tau$ compares compatibility with the corresponding NDCG measure.
}
\vspace*{-\baselineskip}
\label{tab:sc}
\end{table}

For Figure~\ref{fig:cqx} we combined the top-5 crowdsourced answers with the
graded judgments.
Instead, we might focus exclusively on the top-5 answers,
recognizing that a searcher will rarely look beyond these results.
Nothing beyond the top-5 counts,
as if the search engine returned nothing after that point.
The set of ideal rankings now consists of a single
element~---~this single ranking of the top-5 answers~---~ or
perhaps a small number of equivalent rankings if crowdsourcing produced ties.

As a minor point,
under these circumstances ideal rankings are no longer indefinite in the
sense of ~\citet{wmz10}.
Under any circumstances, RBO always leaves a ``residual'',
since rankings cannot practically be computed to infinity.
This residual becomes vanishingly small as rankings become deeper.
However, when $k$ is small this residual can be noticeably large,
and if we limit ideal rankings to just the top-$k$,
then they are not even theoretically indefinite.
As a result,
in this circumstance we apply a normalization for RBO,
as follows:
\begin{equation}
\mbox{NRBO}(R, I) = \frac{\mbox{RBO}(R, I)}{\mbox{RBO}(I, I)}.
\end{equation}
Unless the ideal ranking is relatively shallow,
$\mbox{RBO}(I, I)$ is close to one, but if not,
this formula provides a simple way to normalize out the residual.

While this normalization scales scores into the range $[0,1]$,
it does not matter from a statistical sense,
since the same constant is applied to every run.
Apart from lower values, plots are identical.
However, if $k$ varies from query to query,
this normalization would allow each query to contribute equally
to the magnitude of the average score.
While we do not vary $k$ in this way for the experiments in this paper,
we can imagine this would be helpful in the case of Web search,
for example, where different values of $k$ might be used for
navigational vs.\ informational queries.

\begin{figure}[t]
\centering
\includegraphics[width=3.3in, keepaspectratio]{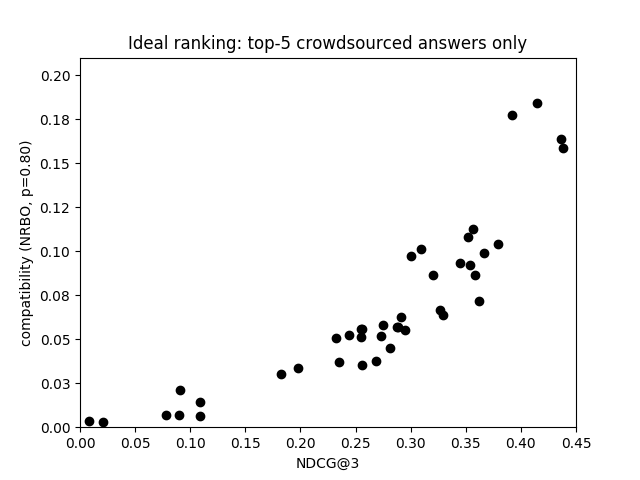}
\caption{
  The relationship between NDCG@3 and compatibility on 
  TREC 2019 Conversational Assistance Track runs
  when ideal rankings are based solely on the crowdsourced top-5 answers.}
\label{fig:cq}
\end{figure}

Figure~\ref{fig:cq} shows the relationship between NDCG@3 and compatibility
when ideal rankings are based solely on crowdsourced top-5 answers.
As shown in Table~\ref{tab:sc} the sensitivity of 73.3\% is lower
than with the combined ideal rankings of Figure~\ref{fig:cqx},
but higher than with graded judgments alone.
The separation between the top-four runs and the rest of the runs remains.

\begin{figure}[t]
\centering
\includegraphics[width=3.3in, keepaspectratio]{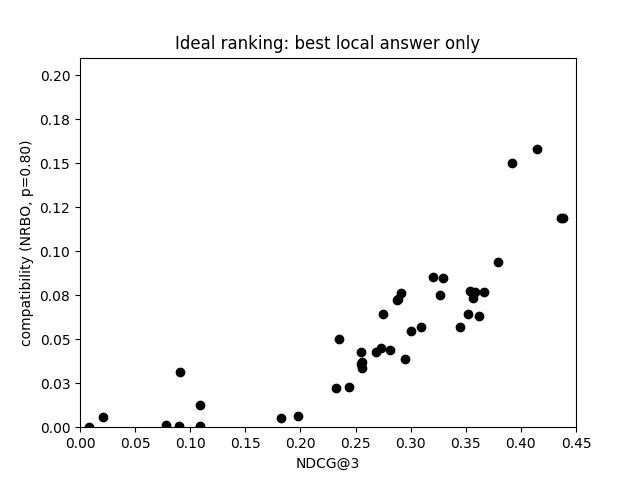}
\caption{
  The relationship between NSCG@3 and compatibility on 
  TREC 2019 Conversational Assistance Track runs
  when ideal rankings are based solely on the single best answer identified
  through dedicated assessment by the research team.}
\label{fig:cq1}
\end{figure}

To go one step further,
Figure~\ref{fig:cq1} shows the relationship between NDCG@3 and compatibility
when ideal rankings are based only on the single best local answer
identified by the research team.
Many runs now have compatibility values close to zero,
even when NDCG@3 values are close to 0.2.
Although sensitivity is now only 55.2\%,
the relative ordering of the top-four runs has not changed.
Using only the single best crowdsourced answer produces a similar result
(not shown).

\section{Conclusion}

It is widely recognized that offline evaluation should focus on the top ranks,
those the searcher will most likely see.
We often report measures of the form \emph{thing}@$k$,
for small values of $k$,
with NDCG@3 providing a typical example.
In effect, these measures evaluate rankers by asking the question:
``What items did the ranker put in the top $k$ ranks?''
In this paper, we turn this question around, asking instead:
``Where did the ranker put the items that should be in the top $k$ ranks?''
By doing this,
we achieve an evaluation measure that is not only focused on the quality of the
top ranked results,
but which is also more sensitive to important differences between rankers.

It is only recently that neural rankers have begun to show significant
improvements over traditional methods on IR tasks~\cite{ylyl19},
and neural methods do not consistently provide the same dramatic improvements
seen on many NLP tasks.
We hypothesize that the lack of dramatic improvement may be due to the
limitations of traditional IR evaluation methodologies,
%with their focus on topical relevance,
which cannot capture important aspects of searcher preferences.
In this paper,
we propose partial preferences focused on the top ranks as a practical
method for capturing these aspects.

While we have demonstrated that our assessment methods can be practically
and affordably applied to an academic evaluation exercise,
we have not as yet applied these methods in a commercial context.
In addition, we have also not explored the cost-benefits tradeoffs
of varying the judging parameters: $k$, $F$, and $P$.
While our current method was kept as simple as possible to make easy for
others to replicate, 
statistical and machine learning methods from the literature might
be extended to partial preferences~\cite{cbch13,ra11,bawgpa13,az14},
reducing assessment effort at the cost of complexity.

In this paper, we piggybacked our work on the existing TREC 2019 CAsT
graded judgments.
Based on our experience,
if top-$k$ partial preferences was the end-goal from the start,
it might be possible to simplify the initial graded assessment to
three grades:
A: ``answers the question''
B: ``provides related information'',
C: ``not relevant''.
The grade-A passages would then become the initial candidate pool,
unless its size is less than $k$,
in which case the grade-B passages would be included.
While this process might produce a larger initial candidate pool and
increase the total number of assessments,
by simplifying the initial graded assessment stage it might speed the overall
process, reducing total costs.
The trade-off depends on the relative cost and consistency of
graded vs.\ preference judgments,
including any savings from reducing the complexity of graded assessment.

\bibliographystyle{ACM-Reference-Format}
\bibliography{paper}

\begin{appendix}

\section{Software and Data Release}
\label{open}

Code and preference judgments are available at
\url{https//github.com/claclark/compatibility}.
Preference judgments are released without personally identifying information,
for which we have University of Waterloo ethics approval.

The implementation of compatibility used for these experiments consists
of a hundred-line Python script,
which is backward compatible with the standard formats used by TREC for
adhoc runs and judgments.
These judgments are expressed as
$(\mbox{\em topic-id}, \mbox{\em document-id}, \mbox{\em preference})$
triples (plus the required but unused ``Q0'' field).

Preferences can be any positive floating point or integer value.
If one document's preference value is greater than another document's
preference value,
it indicates that the first document is preferred over the second.
If preferences are tied,
it indicates that the two documents belong to the same effectiveness level.
The number of effectiveness levels for a topic is defined by the number of
distinct preference values for that topic, and can vary from topic to topic.
In this way, the program can be used directly with many existing TREC runs
and qrels and extended by adding additional preference values.

By default the code computes NRBO,
since this normalization is close to one unless the number of qrels is small.
By default, we report $p = 0.95$, which provides a close match to NDCG@20,
a primary measure for the older TREC Web Tracks.
Overall the code should work ``out of the box'' for typical TREC tasks.

\end{appendix}

\end{document}